\documentclass[aps,twocolumn,showpacs]{revtex4}
\usepackage{graphicx}
\def\bsigma{\mbox{\boldmath $\sigma$}}

\def\ss{\scriptscriptstyle }
\begin{document}
\title{Photoconductivity of an intrinsic graphene}
\author{F.T. Vasko$^{1,2}$}
\email{ftvasko@yahoo.com}
\author{V. Ryzhii$^{2,3}$}
\affiliation{$^{1}$~Institute of Semiconductor Physics, NAS of Ukraine,
Pr. Nauki 41, Kiev, 03028, Ukraine \\
$^{2}$~University of Aizu, Ikki-machi, Aizu-Wakamatsu 965-8580, Japan \\
$^{3}$~Japan Science and Technology Agency, CREST, Tokyo 107-0075, Japan}
\date{\today}

\begin{abstract}
We examine the photoconductivity of an intrinsic graphene associated with  far- 
and mid-infrared irradiation at low temperatures. The model under consideration 
accounts for the excitation of the electron-hole pairs by incident radiation,
the interband generation-recombination transitions due to thermal radiation,
and the intraband energy relaxation due to acoustic phonon scattering. 
The momentum relaxation is assumed to be caused by elastic scattering. 
The pertinent collision integrals are adapted for the case of the massless energy 
spectrum of carriers that interact with the longitudinal acoustic mode 
and the thermal radiation. It is found that the photoconductivity  is determined 
by interplay between weak energy relaxation and generation-recombination processes.
Due to this the threshold of nonlinear response is fairly low.
\end{abstract}


\maketitle

\section{Introduction}

There are two reasons for unusual transport properties of graphene (see  Refs.
~\cite{1,2}): the neutrinolike dynamics of carriers, which is described by 
the Weyl-Wallace model, \cite{3} and the specific features of scattering processes. 
In particular, the high efficiency of interband optical transitions is
associated with a high value of the velocity  $v_{\ss W}\simeq 10^8$ cm/s 
characterizing the linear dispersion relations for the graphene valence and 
conduction bands. This is because the matrix element of interband transitions is 
proportional to $v_{\ss W}$. On the other hand, the coupling of carriers to acoustic 
phonons appears to be weak. As a result, non-equilibrium distributions  of photoexcited 
carriers can readily be realized for the energies smaller than the optical 
phonon energy in the low-temperature region. Due to this, a photoresponse of 
graphene to far- and mid-infrared (IR) irradiations should be strong and a low 
threshold of nonlinear response takes place. To the best of our knowledge, the 
effects of non-equilibrium carriers under far- or mid-IR excitations is not considered 
before: both experimental and theoretical studies of the transport phenomena in 
graphene are performed at weak dc electric fields (see \cite{4} and \cite{5,6VR} 
respectively) or under optical excitation (see \cite{6opt} and references therein).

In this paper, we study the far- and mid-IR photoconductivities of an intrinsic graphene  
at low temperatures. As known, \cite{4,5,6VR} the intrinsic graphene exhibits a maximum 
of the dark resistance. Hence, the effect of photoconductivity of such a material should 
be fairly strong. Since the concentration of carriers in the intrinsic graphene at low 
temperatures is rather small (less $10^{10}$ cm$^{-2}$ at 100 K and lower), one can disregard 
the intra- and interband Coulomb scattering processes. Thus, considering the photoresponse 
of the intrinsic graphene at low-temperatures  one needs to take into account the following 
mechanisms [see Fig. 1(a)]:\\ 
(1) the far- or mid-IR interband photoexcitation;\\ 
(2) the generation-recombination processes due to the interband transitions caused by 
thermal radiation;\\ 
(3) the intraband quasi-elastic scattering on acoustic phonons; \\ 
(4) the scattering  due to a static disorder which is an essential mechanism of the 
momentum relaxation. \cite{2,4,5,6VR} 

\begin{figure}[ht]
\begin{center}
\includegraphics{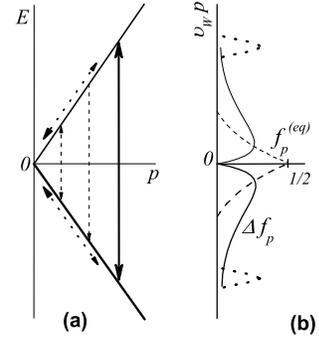}
\end{center}
\addvspace{-1.5 cm}
\caption{(a) Schematic view of the interband and intraband transitions under consideration.
(b) The distribution functions of thermalized and photoexcited carriers, $f_p^{\ss (eq)}$ 
and $\Delta f_p$ (dashed and solid curves, respectively), in the intrinsic graphene at low 
temperatures. Dotted peaks show photoexcited carriers distribution.}
\end{figure}

The shape of the non-equilibrium energy distribution of carriers is determined by 
interplay between the {\it radiative generation-recombination} processes and the 
{\it quasi-elastic energy relaxation}. Indeed, the corresponding relaxation rates,
as we will show below, are of the same order at the  temperatures about 100 K and lower. 
Since both rates are proportional to the density of states and the radiative
transitions are temperature-independent, the generation-recombination processes
not only determine the carrier concentration but also affect the carrier energy 
distribution. Due to an effective recombination of low-energy carriers, their 
energy distribution might become nonmonotonic exhibiting a peak as shown in
Fig. 1(b). Here we restrict ourselves to the linear (with respect to pumping 
intensity) response. The photoconductivity appears to be proportional to the 
concentration of the photogenerated carriers for the case of short-range
momentum relaxation. This concentration as well as the photoconductivity decrease 
with the temperature and the energy of the photoexcitation. However, these 
dependencies are different if the long-range momentum relaxation is essential.

The paper is organized as follows. The basic equations governing the photoconductivity 
in graphene are presented in Sec. II. 
In Sec. III,  we consider the carrier distribution for the case of linear (with respect 
to the photoexcitation rate) response. The results of calculations of the
photoconductivity in this case, including numerical estimates, are presented in Sec. IV. 
The concluding remarks and discussion of the assumptions used are given in Sec.~V. 
In Appendix, the collision integrals for the interband relaxation associated with
thermal radiation and for the intraband relaxation caused by acoustic phonons are 
derived. 

\section{Basic equations}

We describe the linear response of photoexcited carriers to the dc electric field ${\bf E}_0$  
using the steady state kinetic equation for the distribution function $F_{l{\bf p}}$: \cite{7mono}
%1
\begin{equation}\label{eq1}
e{\bf E_0}\cdot\frac{\partial F_{l{\bf p}}}{\partial {\bf p}}= \sum_jJ_j\{F|l{\bf p}\}
+  G\{F|l{\bf p}\}.
\end{equation}
Here $l$ corresponds to conduction ($l$=+1) or valence ($l$=-1) band, ${\bf p}$ is the 2D momentum, 
$J_j\{F|l{\bf p}\}$ is the collision integral for the $j$th scattering mechanism,
($j=D,~LA$, and $R$ correspond to the static disorder, the acoustic phonon scattering,
or the radiative-induced interband transitions, respectively), and $G\{F|l{\bf p}\}$ is the 
interband photogeneration rate. Below we restrict ourselves by the weak dc field case, so that
$F_{l{\bf p}}\simeq F_{lp}+\Delta F_{l{\bf p}}$, where $F_{lp}$ is the symmetric part of the 
distribution function under the photoexcitation and $\Delta F_{l{\bf p}}\propto E_0$ is the
asymmetric addition. We evaluate the in-plane isotropic generation rate, present the kinetic 
equation that governs the distributions functions $F_{lp}$, and discuss the expression for the 
conductivity considered in \cite{6VR}.
   
\subsection{Interband photoexcitation rate}
Within the framework of the Weyl-Wallace model, the carrier dynamics under the in-plane 
ac electric field ${\bf E}_{\omega}\exp (-i\omega t)+ c.c.$
with the frequency $\omega$ is described by the zero-field Hamiltonian
$\hat{h}_{\ss W}=v_{\ss W}(\hat{\bsigma}\cdot{\bf p})$ with the pseudospin Pauli matrix 
$\hat{\bsigma}$ and the harmonic perturbation operator:
%2
\begin{equation}\label{eq2}
\frac{iev_{\ss W}}{\omega}(\hat{\bsigma}\cdot{\bf E}_{\omega})
e^{-i\omega t}+c.c. ~ .
\end{equation}
Following the general consideration of interband photoexcitation (see Sec.~53 in 
Ref. 8), one can obtain the interband photogeneration rate into the 
$l{\bf p}$-state in the form:
%3
\begin{eqnarray}
G\{F|l{\bf p}\} =\frac{2\pi }{\hbar}\biggl( \frac{ev_{\ss W}}{\omega}\biggr)^2
\sum_{l^{\prime}{\bf p}^{\prime}}|(l{\bf p}|\hat{\bsigma}{\bf \cdot E}_{\omega}
|l^{\prime }{\bf p}^{\prime})|^2( F_{l ^{\prime }{\bf p}^{\prime}}- F_ {l{\bf p}}) 
\nonumber \\
\times [ \delta_{\gamma} \left( \varepsilon_{l{\bf p}} -
\varepsilon _{l ^{\prime}{\bf p}^{\prime}}-\hbar \omega \right) + 
\delta_{\gamma}(\varepsilon_{l{\bf p}} -\varepsilon _{l ^{\prime}{\bf p}^{\prime}}
+\hbar \omega)] , ~~~~
\end{eqnarray}
where the energy conservation law has been introduced via the function $\delta_{\gamma} 
(\varepsilon)$ with the phenomenological broadening energy $\gamma$.~\cite{8broad} 

Next, we neglect the in-plane anisotropy of $F_{l{\bf p}}$ assuming 
that the dc electric field ${\bf E_0}$ is sufficiently weak, so that one can 
perform the averaging of the matrix element in Eq.~(3) 
over the in-plane angle:~\cite{9eg}
$\overline{\left|\langle\pm 1,{\bf p}|\hat{\sigma}_{x,y}|\mp 1,{\bf p}\rangle \right|^2 } 
= 1/2$. As a result, the generation rate in $l$th band takes form:
%4
\begin{eqnarray}\label{eq4}
G\{F|l{\bf p}\} =\frac \pi \hbar \left( \frac{ev_{\ss W}E_{\omega}}\omega\right)^2
\left( F_{-lp} - F_{lp}\right) \nonumber \\
\times\delta_{\gamma}\left( \varepsilon_{lp}-\varepsilon_{-lp}-l\hbar\omega \right) 
\end{eqnarray}
with $G\{F_{+l{\bf p}}\} +  G\{F_{-l{\bf p}}\} =  0$ 
due to the  conservation of the net number of the carriers. It is convenient to use in the 
following the electron-hole representation that introduces the electron ($e$) and hole ($h$) 
distribution functions according to the standard replacements \cite{10}:
%5
\begin{equation}\label{eq5}
F_{+1,{\bf p}}\rightarrow f_{e{\bf p}}, \qquad
1-F_{-1,{\bf p}t}\rightarrow f_{h{\bf p}}.
\end{equation}
Considering this substitution, we obtain $G\{F|+l{\bf p}\} 
\rightarrow G\{ f|p\}$ and $G\{ F|-1,p\} \rightarrow -G\{ f|p\}$ with the same generation rate:
%6
\begin{equation}\label{eq6}
G\{ f|p\} =\frac \pi \hbar \left( \frac{ev_{\ss W}E_{\omega}}{\omega}\right)^2\left( 
1-f_{ep}-f_{hp}\right)\delta_{\gamma} \left( 2v_{\ss W}p-\hbar\omega \right),
\end{equation}  
which is symmetric with respect to the electron-hole replacement, 
$f_{ep}\leftrightarrow f_{hp}$.

Considering the case of relatively weak excitation, we assume in the following  
that $f_{e,hp}\ll 1$ in the vicinity of $p_{\omega}\equiv \hbar\omega /2v_{\ss W}$. 
For such a case, $G(f|p)\simeq G_p$ and $G_p$ are  independent of  the carrier distribution: 
%7
\begin{equation}
G_p=\nu_{\ss E}\Delta\left(\frac{p-p_{\omega}}{\delta p}\right), ~~~~\nu_{\ss E}=\frac{\pi}
{\hbar\gamma} \left( \frac{ev_{\ss W}E}\omega \right)^2  .
\end{equation}
Here $\Delta (E)\equiv\gamma\delta_{\gamma}(E)$, $\delta p\equiv\gamma /2v_{\ss W}$ 
and $\nu_{\ss E}$ is the photoexcitation frequency. By disregarding the intraband 
photoexcitation of electrons and holes, i.e. the Drude mechanism of photoexcitation, 
we suppose that  $\delta p\ll p_{\omega}$ or $\gamma\ll\hbar\omega$.

\subsection{Kinetic equation for $f_p$}

Since the collision integrals [see Eqs.~(A.8) and (A.12) in Appendix A] preserve 
their form when $f_{ep}$ is replaced by $f_{hp}$, the electron and hole distributions 
in the intrinsic material under the  photoexcitation are identical (see Fig. 1). In 
addition, the elastic scattering does not affect the symmetric distribution due to the 
energy conservation. As a result, the kinetic equation for $f_{ep}=f_{hp}\equiv f_p$ 
takes form:
%8
\begin{equation}\label{eq5}
J_{\ss LA} \{ f|p\} + J_{\ss R} \{ f|p\} + G_p=0 .
\end{equation}
Here the terms $J_{\ss LA}$ and $J_{\ss R}$ describe the relaxation of electrons (holes) 
caused by the phonon and photon thermostats with the lattice temperature $T$. These 
terms are calculated in Appendix. The term associated with the interband contribution 
$J_{\ss R}\{ f|p\}$ can be presented as
%9
\begin{eqnarray}
J_{\ss R} \{ f|p\} =\nu_p ^{\ss (-)}(1 - f_{p})^2-\nu_p^{\ss ( + )}f_{p}^2 , \\
\nu_p^{(\pm )}=\nu_p^{\ss (R)}\left( N_{2p/p_{\ss T}} + 1/2 \pm 1/2 \right) , \nonumber
\end{eqnarray}
where the generation (or recombination) rate, $\nu_p^{\ss (+)}$ (or $\nu_p^{\ss (-)}$),
is expressed via the Planck function $N_{2p/p_{\ss T}}=[\exp (2p/p_{\ss T})-1]^{-1}$,
where $p_{\ss T}=T/v_{\ss W}$ is the characteristic thermal momentum. The rate of 
spontaneous radiative transitions, $\nu_p^{\ss (R)}$, is presented [see Eqs. (A.5) and 
(A.6) in Appendix] as follows:
%10
\begin{equation} 
\nu_p^{\ss (R)}=\frac{e^2\sqrt\epsilon}{\hbar c}\left(\frac{v_{\ss W}}{c} 
\right)^2 \frac{8v_{\ss W}p}{3\hbar}\equiv\frac{v_rp}{\hbar} 
\end{equation}
where $\epsilon$ is the dielectric permittivity. Here, we have introduced the characteristic
radiative velocity $v_r$.

To derive the expression for the in-plane isotropic collision integral $J_{\ss LA}\{ f|p\}$ 
from (A.12), it is convenient to transform the transition probabilities (A.11) into 
$\overline{W}_{{\bf pp}'}=(W_{{\bf pp}'}+W_{{\bf p}'{\bf p}})/2$ and $\Delta 
W_{{\bf pp}'}=W_{{\bf pp'}}-W_{{\bf p'p}}$, so that $J_{\ss LA}\{ f|{\bf p}\}$ can be 
presented as:
%11
\begin{eqnarray}
J_{\ss LA}\{ f|{\bf p}\} =\sum_{{\bf p'}} \overline{W}_{\bf pp'}\left( f_{\bf p'}-f_{\bf p}
\right) ~~~~~ \\
-\frac{1}{2}\sum_{{\bf p'}}\Delta W_{\bf pp'}\left[ (1-f_{\bf p})f_{\bf p'}+ 
(1-f_{\bf p'})f_{\bf p}\right] . \nonumber
\end{eqnarray}
The energy transfer described by Eq. (A.11) is small because $s\ll v_{\ss W}$ where $s$ is 
the sound velocity. By using also the in-plane averaging $\overline{\Psi (\widehat{
{\bf p},{\bf p}'})|{\bf p} - {\bf p}'|^2}=(p^2+p'^2-pp')/2$, where the overlap factor 
$\Psi\left(\widehat{{\bf p'},{\bf p}}\right)$ is given by Eq. (A.10), one obtains the 
transition probabilities:
%12
\begin{equation}
\left| {\begin{array}{*{20}c} {\overline{W}_{{\bf pp}'} }  \\  {\Delta W_{{\bf pp}'} }  \\
\end{array}} \right|\simeq\frac{{\pi D^2 }}{{\hbar \rho_s L^2 }}\frac{{p^2+p'^2-pp'}}
{2v_{\ss W}^2}\left| {\begin{array}{*{20}c}
   {p_{\ss T}\frac{d^2}{dp'^2}\delta (p - p')}  \\ \\
   {2\frac{d}{{dp'}}\delta (p - p')}  \\
\end{array}} \right| ,
\end{equation}
where $D$ is the deformation potential, $\rho_s$ is the sheet density of 
graphene and $L^2$ is the normalization area. Substituting Eq.~(12) into Eq.~(11) and 
integrating by parts give us the collision integral in the Fokker-Planck form 
\cite{7mono,11pk}:
%13
\begin{eqnarray}
J_{\ss LA}\{ f|p\} =\frac{\nu_p^{\ss (qe)}}{p^2}\frac{d}{dp}\left\{ p^4\left[\frac{df_p} 
{dp}+\frac{f_p (1 - f_p )}{p_{\ss T}}\right]\right\} ,\nonumber \\
\nu_p^{\ss (qe)}=\left(\frac{s}{v_{\ss W}}\right)^2\frac{v_{ac}p}{\hbar} , ~~~ v_{ac}=
\frac{D^2T}{4\hbar^2\rho_sv_{\ss W}s^2} . ~~~
\end{eqnarray}
Here, we have introduced the rate of energy relaxation, $\nu_p^{\ss (qe)}$, and the 
characteristic velocity $v_{ac}$ [compare to Eq. (10)]. 

Since the collision integral (13) is proportional to $p^{-1}d\{\ldots\} /dp$, the 
integration of the kinetic equation (8) over the $p$-axis yields the following 
normalization condition:
%14
\begin{equation}
\int_0^{\infty}dp\rho_pJ_{\ss R}\{ f|p\} +\rho_{p_{\omega}}\nu_{\ss E}\delta p= 0 ,
\end{equation}
where $\rho_p=2p/\pi\hbar^2v_{\ss W}$ is the density of states in graphene. 
Deriving Eq. (14), we have used $\int_0^{\infty}dp\rho_p\Delta [(p-p_{\omega}) /
\delta p] \simeq \rho_{p_{\omega}}\delta p$ if $\delta p\ll p_{\omega}$.

\subsection{Conductivity}
The asymmetric parts of electron and hole distribution functions have similar form, 
and $\Delta f_{e\mathbf{p}}=\Delta f_{h\mathbf{p}}\equiv\Delta f_{\mathbf{p}}$ is given by 
%15
\begin{equation}
\Delta f_{\mathbf{p}}=-\frac{(e\mathbf{E}\cdot \mathbf{p})}p\tau
_p^{\ss (m)}\left( -\frac{df_p}{dp}\right).
\end{equation}
Here $\left[\tau_p^{\ss (m)} \right]^{-1}=v_dp\Psi (pl_c/\hbar )/\hbar$ is the momentum 
relaxation time, where $v_d$ and $\Psi (pl_c/\hbar )$ characterize the strength of 
disorder and the quenching of the long-range scattering. If a random potential, 
$U_{\bf x}$, is characterized by the correlation function $\langle U_{\bf x}U_{\bf x'}\rangle
\equiv\overline{U_d}^2\exp\{-[({\bf x}-{\bf x'})/l_c]^2\}$ with the averaged energy 
$\overline{U_d}$ and the correlation length $l_c$, \cite{6VR} one obtains 
%16
\begin{equation}
v_d=\frac{\pi\overline{U}_d^2l_c^2}{4\hbar ^2v_{\ss W}}, ~~~~
\Psi (z)=\frac{e^{-z^2}}{z^2}I_1\left( z^2\right) ,
\end{equation}
where $I_1(x)$ is the first-order Bessel function of an imaginary argument. 

The conductivity, $\sigma$, is determined by the standard formula \cite{10}
%17
\begin{equation}
\sigma =\frac{4e^2v_{\ss W}}{L^2}\sum_{\mathbf{p}}\tau _p^{(m)}\left( -%
\frac{df_p}{dp}\right) ,
\end{equation}
where the nonequilibrium distribution $f_p$ is determined by Eq. (8). Using the 
relaxation time $\tau _p^{(m)}$ introduced by Eqs. (15) and (16) and integrating by
parts in Eq. (17), the conductivity can be presented as follows:
%18
\begin{equation}
\sigma =\frac{e^2}{\pi\hbar}\frac{2v_{\ss W}}{v_d}\left[2f_{p=0}+\frac{l_c}{\hbar}
\int\limits_0^{\infty}dpf_p\Phi\left(\frac{pl_c}{\hbar} \right)\right] .
\end{equation}
Here, we have introduced the function $\Phi (z)=-\Psi '(z)/\Psi (z)^2$ and take into 
account $\Psi (0)=1/2$. In the case of short-range scattering, $p_{\ss T}l_c/\hbar\ll 1$, 
for the equilibrium conductivity, which corresponds to the distribution $f_p^{\ss (eq)}=
[\exp (p/p_{\ss T})+1]^{-1}$, one obtains $\sigma_{eq}\simeq (e^2/\pi\hbar )
2v_{\ss W}/v_d$. 

\section{Nonequilibrium distribution}
Collecting Eqs. (7-9) and (13) together, we arrive at the equation for the 
symmetric distribution function $f_p$:
%19
\begin{eqnarray}
\frac{\nu_p^{\ss (qe)}}{p^2}\frac{d}{{dp}}p^4 \left[\frac{df_p}{dp}+\frac{f_p 
(1 - f_p )}{p_{\ss T}} \right] +\nu_p^{\ss (\ss R)}\left[ N_{2p/p_T} 
(1 - f_p )^2 \right. \nonumber \\
\left. -(N_{2p/p_T }+1)f_p ^2\right] + \nu_{\ss E} \Delta 
\left( {\frac{p - p_\omega}{{\delta p}}} \right) = 0 , ~~~~~
\end{eqnarray}
which should be solved with the zero-flow boundary condition, $[df_p/dp+f_p(1-f_p)
/p_{\ss T}]_{p\rightarrow\infty}=0$, and the normalization condition (14). Since 
the interband photogeneration is centered in the narrow region $|p-p_\omega |
\leq\delta p$, one can integrate Eq. (19) over this region. Neglecting an exponentially 
weak flow at $p>p_{\omega}$, we consider below the uniform equation  (19), without the
generation term proportional to $\nu_{\ss E}$. Instead of the generation contribution,
we use the boundary condition:
%20
\begin{equation}
\left[\frac{df_p}{dp}+\frac{f_p(1-f_p)}{p_{\ss T}} \right]_{p=p_{\omega}}=
\frac{\nu_{\ss E} \delta p}{\nu_{p_{\omega}}^{\ss (qe)}p_{\omega}^2}  
\end{equation}
At low pumping, when $f_p\simeq f_p^{(eq)}+\Delta f_p$ and $\Delta f_p\ll 1$ 
($\Delta f_p\propto \nu_{\ss E}$), Eqs. (19) and (20) can be presented as
%21
\begin{eqnarray}
\frac{d}{{dp}}p^4\left[ {\frac{{d\Delta f_p }}{{dp}} + \frac{{\Delta f_p }}{{p_{\ss T}}}
\tanh \left( {\frac{p}{{2p_{\ss T}}}} \right)} \right] \nonumber \\
- \frac{{v_r p^2 }}{{v_{qe} \sinh (p/p_T )}}\Delta f_p  = 0 , \\ 
\left(\frac{d\Delta f_p}{{dp}} + \frac{\Delta f_p}{p_{\ss T}} \right)_{p=p_\omega  }  
= \frac{{\nu_{\ss E}\delta p}}{{\nu _{p_\omega  }^{(qe)} p_\omega  ^2 }} . \nonumber
\end{eqnarray}
Simultaneously, the normalization condition (14) takes form:
%22
\begin{equation}
\int_0^{p_\omega}dp\frac{dpp\nu_p^{\ss (R)}}{\cosh (p/p_{\ss T})}\Delta f_p
=\nu_{\ss E}\delta pp_\omega .
\end{equation}
In the region $p\gg p_{\ss T}$, where $\tanh (p/2p_{\ss T})\approx 1$ and $\sinh (p/p_{\ss T}) 
\approx e^{p/p_{\ss T}}/2$, one can neglect $(d\Delta f_p /dp)$ in comparison to
$\Delta f_p /p_{\ss T}$. Consequently, a slow tail of the energy distribution can 
be governed by the first order equation obtained from Eq. (21). The boundary condition 
assumes the form $\Delta f_{p=p_\omega}=\Delta f_\omega$ with 
%23
\begin{equation}
\Delta f_\omega \simeq\frac{\nu_{\ss E}\delta pp_{\ss T}}{\nu_{p_{\omega}}^{\ss (qe)}
p_{\omega}^2} .
\end{equation}
By introducing a connection point $p_c$ according to $p_{\ss T}\ll p_c\ll p_\omega$, we
obtain the following slow-varying solution for the interval $p_c<p<p_\omega$:
%24
\begin{eqnarray}
\frac{\Delta f_p}{\Delta f_{\omega}}\simeq\left( \frac{p_\omega}{p}\right)^4 \exp 
\left( -\frac{{2p_{\ss T}v_r }}{{v_{qe} }}\int\limits_p^{p_\omega}\frac{dp_1}
{p_1^2}e^{-p_1/p_{\ss T}} \right) ~~~ \\ 
=\left( {\frac{{p_\omega  }}{p}} \right)^4 \exp \left[ {-\frac{{2v_r }}
{{v_{qe} }}\left( {\frac{{p_{\ss T}}}{p}} \right)^2 \left( {e^{ - p/p_{\ss T}}-
e^{-p_\omega /p_{\ss T}}}\right)} \right] . \nonumber
\end{eqnarray}

For the low-energy region, $p<p_c$, we search a solution in the form
%25
\begin{eqnarray}
\Delta f_p=\exp\left[ { - \frac{1}{2}\int {dp\left( {\frac{4}{p}+\frac{\tanh (p/2p_T )}
{p_{\ss T}}} \right)} } \right]\delta f_p  \nonumber \\
\approx \frac{{(p_{\ss T}/p)^2 }}{\cosh (p/2p_{\ss T})}\delta f_p , ~~~~~~ 
\end{eqnarray}
where $\delta f_p$ is governed by the second order equation
%26
\begin{equation}
\frac{{d^2\delta f_p }}{{dp^2 }}-\frac{{v_r /v_{qe}}}{p^2\sinh (p/p_{\ss T})}
\delta f_p = 0
\end{equation}
with the parameter $v_r/v_{qe}\gg 1$. Within the WKB approximation \cite{12wkb},
we obtain the solution of Eq. (26) in the form:
%27
\begin{equation}
\delta f_p\simeq C\exp\left[ - \sqrt {\frac{v_r}{v_{qe}}} \int\limits_p^{p_c } 
{\frac{dp_1}{p_1 \sqrt{\sinh (p_1 /p_{\ss T})} }} \right] ,
\end{equation}
where the normalization constant $C$ is determined from the continuity condition
at $p=p_c$ which is transformed into $(p_{\ss T}/p_c)^2\delta f_{p_c}=\Delta f_{p_c}
\cosh (p_c/2p_{\ss T})$. 

As a result, the photoexcited distribution takes the form:
%28
\begin{eqnarray}  
\frac{\Delta f_p}{\Delta f_\omega}\approx\left\{\begin{array}{l} \left(\frac{p_\omega}{p}
\right)^4\exp\left[ -\frac{2v_r}{v_{qe}}\left(\frac{p_{\ss T}}{p}
\right)^2\left( e^{-p/p_{\ss T}}-e^{-p_\omega /p_{\ss T}}\right)\right] , \\ 
~~~~~~~~~~  p_c<p<p_\omega , \\
~~~~~ \{ A_cp_\omega^4 /[2p_c^2p^2\cosh (p/2p_{\ss T})] \} \\
\times\exp\left[ - \sqrt {\frac{v_r}{v_{qe}}} \int\limits_p^{p_c } 
{\frac{dp_1}{p_1 \sqrt{\sinh (p_1 /p_{\ss T})} }} \right] , \\ 
~~~~~~~~~~ 0<p<p_c , \end{array} \right. 
\end{eqnarray}
where
%29
\begin{equation}
A_c\simeq\exp \left[\frac{p_c}{2p_{\ss T}}-\frac{2v_r}{v_{qe}}\left(\frac{p_{\ss T}}{p_c}
\right)^2 e^{-p_c/p_{\ss T}}\right] .
\end{equation} 
The connection point $p_c$ in the distribution (28) is determined by 
%30
\begin{eqnarray}  
\frac{v_{qe}}{v_r}=\frac{A_c}{2}\left(\frac{p_{\ss T}}{p_c}\right)^2
\int\limits_0^{p_c/p_{\ss T}}\frac{dx}{\sinh x\cosh (x/2)} \\
\times\exp \left( -\sqrt{\frac{v_r}{v_{qe}}}\int\limits_x^{p_c/p_{\ss T}} 
\frac{dx_1}{x_1 \sqrt{\sinh x_1}} \right) \nonumber  \\
+2\int\limits_{p_c/p_{\ss T}}^{p_\omega /p_{\ss T}}\frac{dx}{x^2}\exp\left(
-\frac{2v_r}{v_{qe}x^2}e^{-x}-x\right)  \nonumber 
\end{eqnarray}
Equation (30) is a consequence of Eq. (22) under the condition $p_\omega/p_{\ss T}\gg 1$.
The numerical solution of Eq. (30) gives the dependency of $p_c/p_{\ss T}$ on $v_r/v_{qe}$
which can be approximated by the relation $p_c/p_{\ss T}\approx 0.7+(v_r/v_{qe})^{0.4}
/1.5$ with an accuracy about 5 \%. 

\begin{figure}[ht]
\begin{center}
\includegraphics{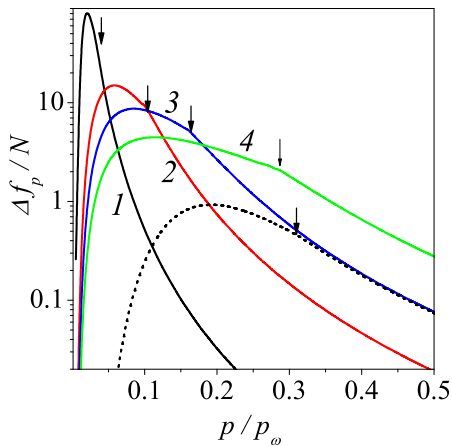}
\end{center}\addvspace{-1 cm}
\caption{(Color online) Distribution function $\Delta f_p$ normalized to 
$N=\Delta f_\omega (p_\omega /p_{\ss T})^2$ under 100 meV photoexcitation at $T=$4.2 K (1),
20 K (2), 40 K (3), and 77 K (4) and under 10 meV photoexcitation at $T=$4.2 K (dotted
curve). The connection points $p_c/p_\omega$ are marked by arrows.}
\end{figure} 
The distribution function $\Delta f_p$, which is normalized by the factor $N=\Delta 
f_\omega (p_\omega /p_{\ss T})^2$ and calculated  for different temperatures and 
$\hbar\omega =$ 100 meV, is  shown in Fig. 2. This function calculated  for $\hbar\omega 
=$ 10 meV at $T=$4.2 K is plotted as well. Due to fairly effective
recombination at $p\sim p_{\ss T}$, the distributions approach zero in the low-energy 
region. On the other hand, the fast decreasing distributions takes place at $p>p_c$. As a
result, the main part of photoexcited carriers is localized in the region below $p_c$
and a peak of distribution grows up as $p_\omega /p_{\ss T}$ increases. In order to
find the conditions of the linear (with respect to $\nu_{\ss E}$) response, we use
$N\propto \omega^{-3}T^{-2}$ and set $\max (\Delta f_p)\simeq 0.3$. Fig. 3 shows
the region of parameters (in the "pumping - temperature" plane at different $\hbar\omega$) 
where the linear response takes place. As seen, the threshold of nonlinear 
response dramatically decreases with decreasing of $\hbar\omega$ and $T$. For instance,
the threshold is about 0.6 $\mu$W/cm$^2$ at $T=$4.2 K and 10 meV excitation.
\begin{figure}[ht]
\begin{center}
\includegraphics{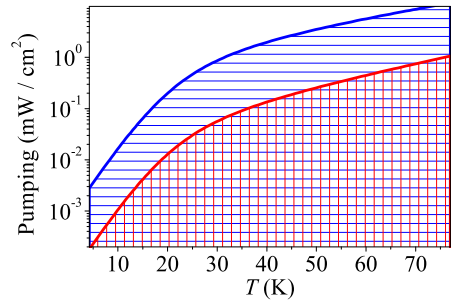}
\end{center}\addvspace{-1 cm}
\caption{(Color online) Region of parameters (shaded by horisontal and vertical lines 
for $\hbar\omega =$ 100 meV and 40 meV, respectively) where the linear regime 
of response takes place.}
\end{figure} 

\section{Photoresponse}
Finally, we consider the photoresponse by using Eq. (18) and distribution 
function (28). The contribution of the photoexcited carriers to the conductivity, 
$\Delta\sigma$, is given by the formula
%31
\begin{eqnarray}
\frac{\Delta\sigma}{\sigma_{eq}}=\Delta f_\omega\Psi\left(\frac{pl_c}{\hbar} \right)^{-1} 
+\frac{l_c}{\hbar}\int\limits_0^{p_\omega}dp\Delta f_p\Phi\left(\frac{pl_c}{\hbar} \right)
\nonumber \\
\simeq \left(\frac{2l_c}{\hbar} \right)^2\int\limits_0^{p_c}dpp\Delta f_p 
+\frac{l_c}{\hbar}\int\limits_{p_c}^{p_\omega}dp\Delta f_p\Phi\left(\frac{pl_c}{\hbar} \right) ,
\end{eqnarray}
where we have neglected $\propto\Delta f_\omega$ contribution in the lower expression.
With the parameter $p_{\omega}l_c/\hbar =\hbar\omega /(2v_{\ss W}\hbar /l_c)$, which 
contains the characteristic energy $2v_{\ss W}\hbar /l_c$=125 meV at $l_c=$10 nm, decreases,
one can use the expansion $\Phi (z)\simeq 4z$. In this case Eq. (31) can be transformed into
the following:
%32
\begin{equation}
\frac{\Delta\sigma}{\sigma_{eq}}\simeq 2\pi l_c^2\Delta n .
\end{equation}
Here $\Delta n=(4/L^2)\sum_{\bf p}\Delta f_p$ is the photoexcited concentration of carriers
which is given by
%33
\begin{equation}
\Delta n\simeq\frac{2}{\pi\hbar^2}\int\limits_0^{p_\omega} dpp\Delta f_p .
\end{equation}
Thus, the photoresponse can be calculated by using the substitution of Eq. (28) into Eq. (31) 
or (33) and numerical integration. 
\begin{figure}[ht]
\begin{center}
\includegraphics{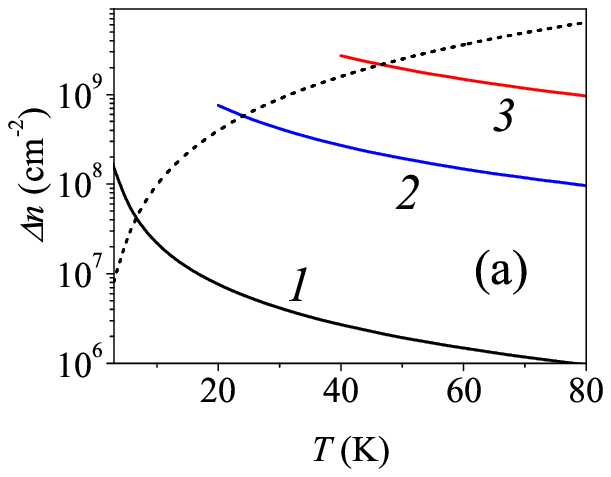}
\includegraphics{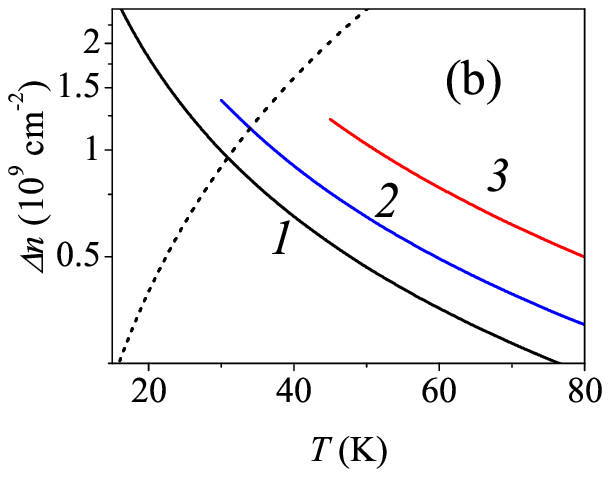}
\end{center}\addvspace{-1 cm}
\caption{(Color online) Photoinduced concentration $\Delta n$ versus temperature: (a) for
the pumping intensities 3 $\mu$W/cm$^2$ (1), 0.3 mW/cm$^2$ (2), and 3 mW/cm$^2$ (3) at
$\hbar\omega$=100 meV and (b) for the excitation energies $\hbar\omega$=140 meV (1), 100 meV 
(2), and 60 meV (3) at the intensity 1 mW/cm$^2$. Dotted curve shows the equilibrium 
concentration.}
\end{figure} 

Consider first the short-range scattering case, $p_\omega l_c/\hbar\ll 1$, which is 
described by the photoinduced concentration, $\Delta n$. The temperature dependencies of 
$\Delta n$, which are calculated for different intensities and different frequencies of excitation,
are shown in Figs. 4(a) and 4(b), respectively. Curves 2 and 3 are plotted for the
linear response region (see Fig. 3). The photoinduced concentration decreases with increasing 
$T$ and $\hbar\omega$, whereas it increases with pumping intensity. Fig. 5 presents the 
spectral dependencies of $\Delta n$ for different temperatures. Despite of $\Delta n$ 
exceeds (or be comparable with) the equilibrium concentration, $n_{eq}\simeq 0.52 (T/\hbar 
v_{\ss W})^2$, the relative photoconductivity (32) does not exceed 10$^{-2}$ for the 
short-range scattering case, when $l_c\leq$10 nm for the parameters used.

\begin{figure}[ht]
\begin{center}
\includegraphics{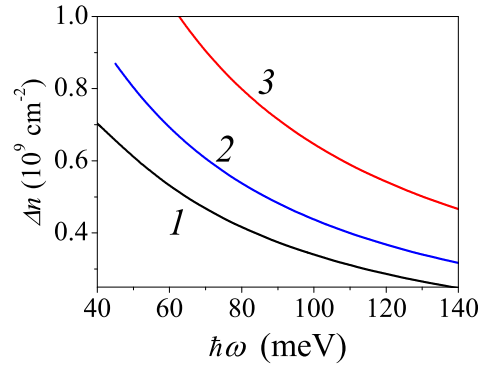}
\end{center}\addvspace{-1 cm}
\caption{(Color online) Spectral dependencies of photoinduced concentration for $T$=50 K 
(1), 65 K (2), and 77 K (3) at the pumping intensity 1 mW/cm$^2$.}
\end{figure} 
\begin{figure}[ht]
\begin{center}
\includegraphics{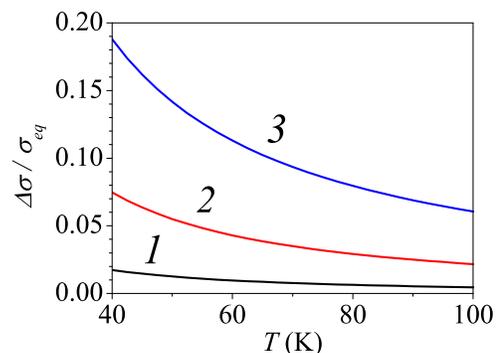}
\end{center}\addvspace{-1.2 cm}
\caption{(Color online) Relative photoconductivity $\Delta\sigma /\sigma_{eq}$ vs temperature
for different correlation lenths $l_c$=10 nm (1), 20 nm (2), and 30 nm (3) at the pumping 
intensity 3 mW/cm$^2$ and $\hbar\omega$=100 meV.}
\end{figure} 

\begin{figure}[ht]
\begin{center}
\includegraphics{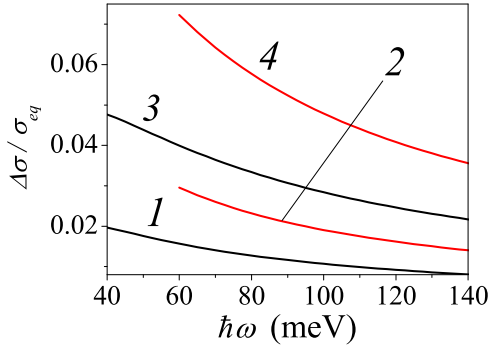}
\end{center}\addvspace{-1.2 cm}
\caption{(Color online) Spectral dependencies of relative photoconductivity for the
pumping intensity 1 mW/cm$^2$ and the correlation lengths $l_c$=20 nm at $T$=77 K (1)
or $T$=50 K (2) and 20 nm at $T$=77 K (3) or $T$=50 K  (4).}
\end{figure} 

In the general case $p_\omega l_c/\hbar\sim 1$, which occurs if $l_c\geq$10 nm, by using 
Eqs. (28) and (31) we obtain $\Delta\sigma /\sigma_{eq}$ vs $T$ and $\hbar\omega$ 
relations as shown in Figs. 6 and 7, respectively. Once again the relative 
photoconductivity decreases with $T$ or $\hbar\omega$; these dependencies are 
similar to $\Delta n$ vs $T$ and $\hbar\omega$ but an additional dependency on $l_c$
becomes essential. In Fig. 7 (curves 2 and 3) we restrict ourselves by the linear 
response region. Since $\Delta\sigma /\sigma_{eq}$ increases with increasing of $l_c$
or intensity, for the above-discussed region of parameters one obtains $\Delta\sigma 
/\sigma_{eq}\sim$ 1.

For the case of far-IR excitation with the energy $\hbar\omega=$10 meV and intensity 
1$\mu$W/cm$^2$, one obtains the photoinduced concentration $\Delta n\simeq 4.1\cdot 10^7$ 
cm$^{-2}$ at 4.2 K while the equilibrium concentration is about $1.7\cdot 10^7$ cm$^{-2}$. 
Since the 
short-range scattering condition is valid now up to $l_c\leq$30 nm, the relative 
photoconductivity $\Delta\sigma /\sigma_{eq}\propto l_c^2$ does not exceed 0.3 \%.

\section{Concluding remarks}
In the present work, we have considered the photoconductivity of intrinsic graphene
under far- or mid-IR excitation of electron-hole pairs. We demonstrated that not only
carrier concentration but also the energy distribution substantially depend on the
parameters of excitation (frequency and pumping intensity) and the temperature. In 
contrast to the customary semiconductor materials, in which the recombination is a most 
slow process, interplay between recombination-generation and energy relaxation in
graphene is crucial. The carrier distribution smeared up to the energy of photoexcitation 
can be realized due to a weak coupling to the phonon thermostat and effective radiative 
interband transitions.

In addition to the physical peculiarities presented, the technical results obtained
might be useful for theoretical descriptions of the nonequilibrium carriers in graphene
(heating under dc field, nonlinear optical properties, etc). We have evaluated the 
generation rate that describes the interband pumping of electron-hole
pairs and described by the generation-recombination processes caused by the interaction 
of carriers with the thermal radiation. We also evaluated the Fokker-Planck collision
integral governed the quasi-elastic energy relaxation due to the deformation interaction 
between carriers and acoustic phonons. 

Let us briefly discuss the assumptions used in our treatment. The main restrictions 
arise from the low-concentration approximation, when one can neglect the intra- and
interband Coulomb scattering. This approach corresponds to an intrinsic graphene 
at low temperatures which is sensitive to photoexcitation due to a high resistance. We
also do not consider a disorder-induced channel of generation-recombination and restrict 
ourselves taking into account the simplest scattering mechanisms (elastic scattering by 
static disorder and deformation interaction with acoustic phonons). These restrictions 
are caused by the lack of data on relaxation process. However, the model used allows us
to describe important peculiarities of photoresponse. Other assumptions are a rather 
standard for the calculations of the transport characteristics. We used the isotropic 
energy spectrum both for carriers and phonons, and consider the quasi-elastic electron-phonon 
scattering. We neglect an interaction with optical phonons because the optical phonon 
energy substantially exceeds $T$ and $\hbar\omega$. These restrictions are valid for 
low temperatures in the far- and mid-IR spectral regions.  

Besides this, we restrict ourselves by the analytical consideration of the limiting
case presented in Sec. III and do not perform a numerical solution of Eq. (19) with
a subsequent integration of (18). This is because the lack of a precise data
concerns both momentum relaxation (see the discussions in Refs. 4-6) and phonon 
scattering (see \cite{phonon} and Refs. therein). Although the analytical consideration
does not provide a complete qualitative description, the relations obtained permit
one to estimate a character of photoresponse for the linear regime with
arbitrary scattering parameters.
 
In closing, the obtained results demonstrated a marked photosensitivity of intrinsic
graphene at low temperatures. This allows one to analyze the relaxation processes in
graphene by using a photoconductivity data. In order to check a potential of graphene-based 
detector applications, one needs to perform a numerical modeling, including the nonlinear 
regime of response.

\appendix*
\section{Collision Integrals}
Below we will evaluate the collision integrals used in Sec. II and our consideration is 
based on the general expression \cite{7mono}:
%A1
\begin{equation}
J\{ F|\alpha\} =\sum_{\alpha '}\left[ {W_{\alpha '\alpha }(1 - F_\alpha  )
F_{\alpha '}- W_{\alpha \alpha '} (1 - F_{\alpha '} )F_\alpha }\right] , 
\end{equation}
where $F_\alpha$ is the distribution function over $\alpha$-state. The transition
probabilities, $W_{\alpha \alpha '}$, are connected by the detailed balance requirement: 
$W_{\alpha '\alpha}=\exp [-(\varepsilon_\alpha -\varepsilon_{\alpha '})/T]$ $\times W_{\alpha
\alpha '}$ and are determined through
%A2
\begin{eqnarray}
W_{\alpha\alpha '}=\frac{2\pi}{\hbar}\sum_q|\langle\alpha '|\hat{\chi}_q|\alpha 
\rangle |^2 ~~~~~~ \\
\times\left[{(N_q  +1)\delta (\varepsilon_\alpha -\varepsilon_\alpha-\hbar\omega_q )
+ N_q \delta (\varepsilon_\alpha -\varepsilon_\alpha + \hbar \omega _q )}
\right] . \nonumber 
\end{eqnarray}
Here $N_q$ is the Planck distribution of the $q$th boson (phonon or photon) mode
of frequency $\omega_q$ and the matrix element $\langle\alpha '|\hat{\chi}_q|\alpha\rangle$ 
describes the electron-boson interaction.

\subsection{Radiative-induced transitions}
First, we consider the interaction with thermal radiation when the secondary-quantized 
radiation field should be substituted into perturbation operator (2) (see Secs. 20 
and 38 in \cite{7mono}). As a result, the operator $\hat{\chi}_q$ in (A.2) takes the form: 
%A3
\begin{equation}
\hat{\chi}_{\eta ,{\ss{\bf Q}}}= v_{\ss W}\sqrt{\frac{2\pi \hbar e^2 }{\omega_{\ss Q}
\epsilon V}}\left( \hat{\bsigma}\cdot {\bf e}_{\eta ,{\ss{\bf Q}}}\right) ,
\end{equation}
where $V$ is the normalization volume, $\omega_{\ss Q}$ and ${\bf e}_{\eta ,
{\ss{\bf Q}}}$ are the frequency and the polarization vector of the photon mode with the 
3D wave vector $\bf Q$ and the polarization $\eta$. We have also neglected in (A3) 
the in-plane momentum transfer under interband transitions. Calculations 
of the matrix elements (A.3) are performed similar to Sec. IIA \cite{9eg} and we obtain
%A4
\begin{equation}
|\left\langle { \pm 1,{\bf p}} \right|\hat{\chi}_{\eta ,{\ss{\bf Q}}} \left| {\mp 1,
{\bf p}} \right\rangle |^2  = v_{\ss W}^2 \frac{{\pi \hbar e^2 }}{{\omega_{\ss Q}\epsilon V}}
|{\bf e}_{\eta ,{\ss{\bf Q}}}^{\ss\|}|^2
\end{equation}
while the intraband transitions are forbidden. The averaging over polarization and 
direction of photons gives $\overline{|{\bf e}_{\eta ,{\ss{\bf Q}}}^{\ss\|}|^2}=2/3$. 

After substituting Eq. (A.4) to general expression (A.2), we obtain the interband 
probabilities:
%A5
\begin{eqnarray}
\left| {\begin{array}{*{20}c} {W_{+1p,-1p}}  \\ {W_{ - 1p, + 1p}}  \\
\end{array}} \right| =\frac{2(2\pi ev_{\ss W})^2}{3\epsilon}
\int\frac{d{\bf Q}}{(2\pi )^3 } \omega_{\ss Q}^{-1} \\
\times\left| {\begin{array}{*{20}c} {N_{\ss Q}+1}  \\  N_{\ss Q}  \\
\end{array}} \right|\delta (2v_{\ss W}p - \hbar\omega_{\ss Q} ) \nonumber
\end{eqnarray}
where $N_{\ss Q}$ stands for the Planck number of photons with temperature 
$T$. Taking the integral over $\bf Q$-space we obtain
%A6
\begin{equation} 
\left| {\begin{array}{*{20}c} {W_{+1p,-1p}}  \\ {W_{-1p,+1p}}  \\
\end{array}} \right| = \nu_p^{\ss (R)} \left| {\begin{array}{l}
 N_{2p/p_{\ss T}} + 1  \\ ~~~~~ N_{2p/p_{\ss T}} \\ \end{array}} \right| ,
\end{equation}
where the rate $\nu_p^{\ss (R)}$ is given by Eq. (10).

By using (A.6), one can transform collision integral (A.1) as follows
%A7
\begin{equation}
J_{\ss R}\{ f|\pm 1,p\} = \nu_p^{\ss (\mp )}(1 - F_{\pm 1,p} )F_{\mp 1,p}  
-\nu_p^{\ss (\pm )}(1 - F_{\mp 1,p} ) F_{\pm1,p} , 
\end{equation}
where the rates of generation ($\nu_p^{\ss (+)}$) and recombination 
($\nu_p^{\ss (-)}$) are given by Eq. (9). We further use the electron-hole 
representation introduced by Eq. (5) and the radiative collision integral for 
electrons takes the form:
%A8
\begin{equation}
J_{\ss R} \{ f|ep\} = \nu_p^{\ss (-)}(1-f_{ep})(1-f_{hp}) 
-\nu_p^{\ss ( + )}f_{ep}f_{hp} 
\end{equation}
and $J_{\ss R}\{ f|hp\} =-J_{\ss R}\{ f|ep\}$, which are in agreement with the particle conservation 
law. Here the recombination term is proportional to $f_{ep}f_{hp}$, while the generation
contribution is proportional to $(1-f_{ep})(1-f_{hp})$, also $\nu_p^{\ss (+)}$ and 
$\nu_p^{\ss (-)}$ mean the rates of spontaneous emission and absorption.

\subsection{Acoustic phonon scattering} 
We further consider the collision integral that is described the intraband 
transitions caused by the acoustic phonon scattering while the interband 
transitions are forbidden due to the condition $s\ll v_{\ss W}$. The main contribution 
to the acoustic phonon scattering appears due to the deformation interaction with 
longitudinal vibrations \cite{ando}, $D\nabla\cdot{\bf u}_{\bf x}$, where 
${\bf u}_{\bf x}$ is the displacement vector of LA-mode. By using the quantized 
displacement operator, one obtains the matrix element of Eq. (A.2) in the standard 
form \cite{10}:
%A9
\begin{equation}
\left|\left\langle l'{\bf p}'|\hat{\chi}_{\bf q}|l{\bf p}\right\rangle \right|^2\simeq 
\delta_{{\bf p}',{\bf p} + \hbar {\bf q}}|C_{|{\bf p}-{\bf p'}|/\hbar}|^2\Psi\left(
\widehat{{\bf p'},{\bf p}}\right) 
\end{equation}
Here, $\bf q$ is the 2D wave vector of phonon with frequency $\omega_q=sq$ and
$|C_q|^2=D^2\hbar\omega_q/(2\rho_s s^2L^2)$ stands as the electron-phonon matrix element. 
The overlap factor, $\Psi\left(\widehat{{\bf p'}, {\bf p}}\right)$, is given 
by \cite{9eg}:
%A10
\begin{equation}
\Psi\left(\widehat{{\bf p'},{\bf p}}\right)\equiv\left|\left\langle l{\bf p}'|l{\bf p}
\right\rangle\right|^2 =\frac{1+\cos\widehat{{\bf p'},{\bf p}}}{2} .
\end{equation}

Since (A.9) does not depend on $l=\pm 1$ (electron-hole symmetry), we obtain
$W_{+1{\bf p},+1{\bf p}'}= W_{-1{\bf p},-1{\bf p'}}=W_{\bf pp'}$ and the 
transition probability is given by
%A11
\begin{eqnarray}
W_{\bf pp'}=\frac{2\pi}{\hbar }\Psi (\widehat{{\bf p},{\bf p'}})\left| 
{C_{|{\bf p}-{\bf p}'| /\hbar } } \right|^2  ~~~~~ \\
\times\left\{ (N_{|{\bf p}-{\bf p'}|/\hbar}+ 1)\delta [v{\ss W}(p - p')-s|{\bf p}
-{\bf p}'|] \right. \nonumber \\
\left. + N_{|{\bf p}-{\bf p}'|/\hbar}\delta [v_{\ss W}(p - p') + s|{\bf p}-{\bf p}'|] 
\right\} , \nonumber
\end{eqnarray}
where $N_q$ is the Planck distribution of phonons with temperature $T$. By using the 
electron-hole representation, see Eq. (5), one transforms collision integral (A.1) into the form:
%A12
\begin{eqnarray}
J_{\ss LA}\{ f|k{\bf p}\} =\sum_{\bf p'}\left[ W_{\bf p'p}(1-f_{k{\bf p}})
f_{k{\bf p'}} \right. \nonumber \\
\left. - W_{\bf pp'} (1-f_{k{\bf p'}})f_{k{\bf p}}\right] , 
\end{eqnarray}
with transition probabilities (A.11) that are the same for electrons ($k=e$) 
and holes ($k=h$).


\begin{thebibliography}{52}
\bibitem{8broad}
One can estimate the broadening $\gamma\sim\hbar /\tau_{p_\omega}^{\ss (m)}\simeq v_{ac}
p_\omega$, so that $\delta p/p_\omega\simeq v_{ac}/v_{\ss W}$.
\bibitem{temp}
The low-temperature restriction on the consideration performed appears due to the
quasielastic approach in the transition probabilities (12) and results in the condition 
$sp_{\omega}\ll T$ which is valid for $T>$4 K if $\hbar\omega =$100 meV.
\end{thebibliography}
\end{document}